\documentclass[submit]{smj}
\usepackage{amsmath}
\usepackage{natbib}
\usepackage{graphicx}
\usepackage{algorithm}
\usepackage{algpseudocode}
\usepackage{amsfonts}

\usepackage{natbib}
\usepackage{xcolor}
\usepackage{url} 
\usepackage{hyperref}

\usepackage{tabularx}
\Author{Martje Rave\Affil{1}, 
        Göran Kauermann\Affil{1}, 
}
\AuthorRunning{Martje Rave \textrm{et al.}}

\Affiliations{

\item Department of Statistics,
             Faculty of Mathematics, Informatics and Statistics,
             Ludwig-Maximilians-Universität München, Germany

}   

\CorrAddress{Martje Rave, 
             Chair of Applied Statistics in Social Sciences, Economics and Business, 
             Department of Statistics,
             Faculty of Mathematics, Informatics and Statistics,
             Ludwig-Maximilians-Universität München,
             Ludwigstr. 33
             80539 München
             Germany}
\CorrEmail{martje.rave@stat.uni-muenchen.de}
\CorrPhone{(+49)\;89\;2180\;2248}
\CorrFax{ (+49)\;89\;2180-5040}

\Title{The Skellam Distribution revisited - Estimating the unobserved incoming and outgoing ICU COVID-19 patients on a regional level in Germany}
\TitleRunning{The Skellam Distribution revisited}

\Abstract{
With the beginning of the COVID-19 pandemic, we realized the need for comprehensive data collection and its provision to scientists and experts for proper data analyses. In Germany, the Robert Koch Institute (RKI) has tried to keep up with this demand for data on COVID-19, but there were (and still are) relevant data missing that are needed to understand the whole picture of the pandemic. In this paper, we  take a closer look at the severity of the course of COVID-19 in Germany, for which ideal information would be the number of incoming patients to  ICU units. This information was (and still is) not available. Instead, the current occupancy of ICU units on the district level was reported daily. We demonstrate how this information can be used to predict the number of incoming as well as released COVID-19 patients using  a stochastic version of the Expectation Maximisation algorithm (SEM). This in turn, allows for estimating  the influence of  district-specific and age-specific  infection rates as well as further covariates, including spatial effects, on the number of incoming patients.   The paper demonstrates that even if relevant data are not recorded or provided officially, statistical modelling allows for reconstructing them. This also includes the quantification of uncertainty which naturally results from the application of the SEM algorithm. 
}


\begin{document}

\maketitle


\section{Introduction}


Albeit its atrocity, in its aftermath, the COVID-19 pandemic has taught Germany, among many other countries, the shortcomings of inadequate data availability in its healthcare system. In fact, in Germany, while intensive care unit (ICU) occupancy at some point was provided by the DIVI (German interdisciplinary association of Intensive and emergency medicine), the important numbers of newly hospitalised patients (incoming) and released patients (outgoing), either cured or deceased,  has (until now) not been included in the database. While this can be criticised  since the most relevant number, which measures the pressure of the disease - the number of incoming patients -  is not officially released, we show in this paper how to disentangle incoming and outgoing patients from pure occupancy data using statistical models. This, in particular, allows us to investigate how hospitalisations depend on time, age  and spatial factors. 

We assume that admission to and release of the ICU units follow Poisson distributions with inhomogeneous intensities. Consequently, the changes in ICU occupancy result from the difference between incoming and outgoing patients. This in turn gives the framework of the Skellam distribution, originally introduced by \citet{skellam1948probability}.  The distribution is easily described as resulting from the difference of two independent Poisson distributed random variables. This distributional approach  has been used in different settings.  For instance, in sports statistics \citet{karlis2009bayesian} apply the distribution for modelling of the goal difference in football games. In network analysis, \citet{gan2018approximation} and
\citet{schneble2020estimation} look at network flows while \citet{koopman2014dynamic} utilise the idea to model  financial trades. Further application areas include image analysis when comparing intensity differences of pixels, see e.g.  \cite{hwang2007sensor}, \cite{hwang2011difference} or \cite{hirakawa2009wavelet}. Extensions  towards bivariate Skellam processes are provided e.g. in  \cite{genest2014bivariate}, see also \cite{aissaoui2017second}. A general discussion on the Skellam distribution and its application fields is provided in \cite{tomy2022retrospective}.
In this paper, we provide an application of the Skellam distribution for disentangling incoming and outgoing patients in ICUs.

The occupancy of ICU units was  a central component of the COVID-19 pandemic. Numerous tools have been developed for forecasting the number of patients who require ICU admission, see e.g. \citet{grasselli2020critical}, \citet{goic2021covid}, \citet{covid2020forecasting} or \citet{farcomeni2021ensemble} to just mention a few. Our focus in this paper is not on primarily on prediction but on investigating the risk of admission and how this depends on the infection rates and further covariates, including spatial components. 

Parameter estimation in the Skellam distribution is cumbersome due to its non-analytic form of the likelihood function. We refer to \cite{lewis2016package} or \cite{aissaoui2017second} who pursue moment-based estimation. In this paper, we show that the stochastic expectation maximisation (SEM) algorithm is a suitable and numerically stable alternative to available estimation routines. Originally proposed by \cite{celeux1996stochastic}, the stochastic version of the EM algorithm gained interest in recent years, in particular in mixture models, see e.g. \cite{noghrehchi2021selecting}.
We also refer to \cite{nielsen2000stochastic}  for asymptotic results on the algorithm. The EM algorithm relates the estimation to a missing data problem, which is easily described. We assume that instead of the complete data with incoming and outgoing patients, we only observe the changes in occupancy of ICUs. In other words, the exact number of incoming and outgoing is missing. Replacing these missing numbers  iteratively with simulated numbers, based on the current estimates of the model, provides the stochastic version of the `E'-step. This, in turn, leads to full data, which allows for standard maximum likelihood estimation of two Poisson processes - the M step. The algorithm is easily implemented, and Rubin's rule, \cite{rubin1976inference}, provides inference statements. 

A particularly interesting attribute that this approach provides is the simplification of the initial complexity of the problem. We are able to break our problem down from a fairly complex distributional assumption, with respect to deriving an association between the infection rates and the number of incoming and outgoing patients, to land at essentially two iteratively updated generalised additive models (GAMs) with simulated responses, each response simultaneously sampled from a joint distribution, comprised of the product of two Poisson distributions. This allows us to not only circumvent rather cumbersome calculations and modifications of the first and second derivative of the Skellam distribution, as, for example, shown by \cite{MarcBessel}, but also almost effortlessly interpret the association between the number of incoming and outgoing patients and the infection rate.

The paper is structured as follows. In Section 2, we give a detailed data description. In Section 3, we elaborate on the model approach to our problem, while in Section 4, we will provide the results of our model approach. A simple simulation exercise to validate our findings can be found in Section 5, and in Section 6, we conclude our paper which also includes a discussion of the shortcomings of our approach.



\section{Data Description}
\begin{figure}[h]
    \centering
    \includegraphics[width=0.9\textwidth]{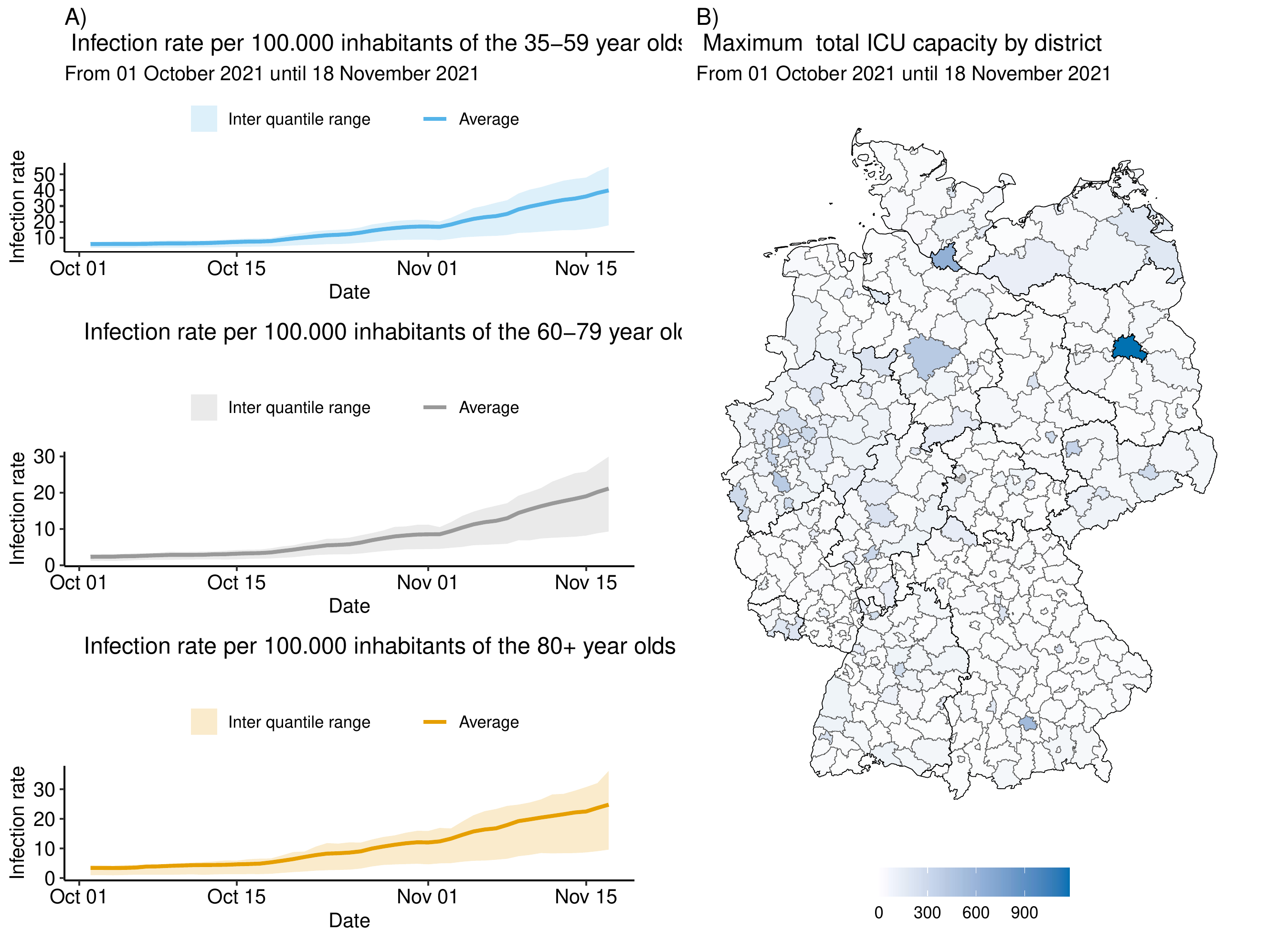}
    \caption{A: Summary over all districts of the infection rate per 100.000 inhabitants by age group, `45-59' year-olds, `60-79' year-olds and `80+' year-olds displayed by date, from the $1^{st}$ of October 2021 until the $18^{th}$ of November 2021, B: The maximum capacity of ICU beds per given district over the time span from the $1^{st}$ of October 2021 until the $18^{th}$ of November 2021 by district}
    \label{fig:Descriptive}
\end{figure}

The database for our analyses consists of two main components; data on COVID-19 infections and data on the ICU occupancy of COVID-19 patients. 
The infections and the ICU occupancy are collected by the German health care departments, recorded by the \citet{Rki2021} (RKI), the German federal government agency and scientific institute responsible for health reporting and disease control, and published 
by the RKI and \citet{Divi2021}, respectively. We here focus on  data during the fourth  infection wave in Germany, i.e. from the $2^{nd}$ of October 2021 until the $17^{th}$ of November 2021, though the method is readily extendable to other time frames. We visualise the average infection rates overall districts in Figure \ref{fig:Descriptive} (left-hand side).


The RKI collects and publishes data on infections on a daily basis. Due to privacy protection, the RKI aggregates the number of COVID-19 patients, ICU and general admission, over NUTS3 districts, \cite{NUTS32021}, and other demographic groups. These, namely, the age categories; `0-4' year-olds, `5-14' year-olds, `15-34' year-olds, `35-59' year-olds, `60-79' year-olds and `80+' year-olds and the sex; `male', `female' and `not disclosed'. The infections are aggregated over the age groups.
The data can be directly downloaded through the \hyperlink{https://www.arcgis.com/sharing/rest/content/items/f10774f1c63e40168479a1feb6c7ca74/data}{ArcGIS website}, \cite{Rki2021}. The infection rates per 100.000 inhabitants are then calculated as a weekly average for each age group. For each district, the infection rate is averaged over the seven days immediately preceding the respective observed day change in ICU occupancy. 




The data on ICU occupancy is also collected by the RKI and published by \hyperlink{https://www.divi.de/register/tagesreport}{DIVI}, \citep{Divi2021}. This data is also on a district level, however, the occupancy can only be differentiated by the number of beds occupied by patients infected with COVID-19, by the number of beds occupied by patients not infected with COVID-19 and the number of empty beds, the sum of which is the overall ICU capacity in a given district on a given date. We here take the COVID-19 patients into account and visualise the ICU data for one day in Figure \ref{fig:Descriptive} (right-hand side). 

Conveniently, both data sets can also be found in the daily updated \hyperlink{https://github.com/robert-koch-institut}{GitHub repository} maintained by the RKI, \citet{GithubRKI}.  


\section{Model}

\subsection{Assumption}

Let $Y_{(t,d)}$ be the number of COVID-19 ICU patients  in a given district $d$ at day $t$. This is the official number issued by DIVI, described above and freely accessible from the given sources. 
We define with $I_{(t,d)}$ the number of incoming patients in district $d$ at day $t$, which is the number of newly admitted COVID-19 patients in the ICUs located in district $d$. Accordingly, we denote with $R_{(t,d)}$ the number of released patients, meaning that they are discharged, deceased or transferred to a non-ICU unit. We assume both to come from an inhomogeneous Poisson process such that 
\begin{align}
\label{eq:in}
I_{(t,d)}  & \sim \mbox{Poisson}\left( \lambda^I_{(t,d)}\right) \\
\label{eq:out}
R_{(t,d)}  & \sim \mbox{Poisson}\left( \lambda^R_{(t,d)} \right).
\end{align} 
The explicit modelling of the intensities $\lambda^I_{(t,d)}$ and $\lambda^R_{(t,d)}$ is of primary interest and discussed in depth later in this section. For now, note that (\ref{eq:out}) is an approximation, and formally we have a right censored Poisson distribution with $R_{(t,d)} \le Y_{(t-1,d)}$ since no   more patients can be released than are currently in the ICU. We can omit this point, though, since, based on the disease, we know that not all patients are discharged at a time, so the formal censoring does not play any practical relevance due to a generally small discharge intensity $\lambda_{(t,d)}$.

With these definitions we can now define the difference $\Delta_{(t,d)}$ in occupancy of COVID-19 ICU patients per  district $d$ and day $t$ to the the previous day $t-1$ through
\begin{equation}
\label{eq:diff}
    \Delta_{(t,d)}=Y_{(t,d)}-Y_{(t-1,d)}=I_{(t,d)}-R_{(t,d)}.
\end{equation}
Assuming independence for  the number of incoming and  outgoing ICU COVID-19 patients together with (\ref{eq:in}) and (\ref{eq:out}) leads to 
a Skellam distribution \cite{skellam1948probability}. 
\begin{align}
\label{eq:skellam}
    \Delta_{(t,d)}\sim \mbox{Skellam} (\lambda^I_{(t,d)}, \lambda^R_{(t,d)}).
\end{align}
Before we derive how to estimate the two intensities in (\ref{eq:skellam}) we want to discuss the suitability of the distributional assumptions. Note that the approach relies on independence of $I_{(t,d)}$ and $R_{(t,d)}$. This would be violated if discharges of the ICU in $t$ depend on the number of incoming patients in $t$.
A  conceivable scenario where $I_{(t,d)}$ and $R_{(t,d)}$ are dependent results if the ICUs get to their limit capacity and triage of patients is inevitable. This situation has not been observed in Germany - over the entire course of the pandemic - so we can argue that assuming independence  between incoming and outgoing patients is reasonable. 

Finally, the intensities $\lambda^I_{(t,d)}$ and $\lambda^R_{(t,d)}$ are modelled to depend on a set of covariates denoted by $\mathbf{x}_{(t,d)}$ as well as previous data. To be specific, we set 
\begin{align}
\label{eq:model-in}
\lambda^I_{(t,d)} & = \exp\left( \eta_{(t,d)}^I + s^I(t) + h^I(\mbox{longitude}_d, \mbox{latitude}_d)\right), \\
\label{eq:model-out}
\lambda^R_{(t,d)} & = \exp\Big( \eta_{(t,d)}^R + s^R(t) + h^R(\mbox{longitude}_d, \mbox{latitude}_d)+ \underbrace{\log(\sum^{j=t}_{j=t-56}\omega_j \hat{I}_{(j,d)})}_{\mbox{= offset}} \Big),
\end{align}
where $\eta_{(t,d)}^I$ and $\eta_{(t,d)}^R$ are the linear combinations of the covariates included in the models. Namely, the logged infection rates of the age groups `$35-59$' year-olds, `$60-79$' year-olds and `$80+$' year-olds, as well as the weekday, included as a categorical variable, with Friday as its reference category. $s^I(t)$ and $s^R(t)$ are smooth functions in time, and $h^I(\mbox{longitude}_d, \mbox{latitude}_d)$ and $h^R(\mbox{longitude}_d, \mbox{latitude}_d)$ are two-dimensional smooth functions over the coordinates of the centroids of the respective districts. Note that $I_{(j,d)}$ is not observed, and we, therefore, replace it with its simulated value from the `E'-step. Moreover, the  weights $\omega_j$ are fixed and not estimated but instead obtained from duration time models for COVID-19 patients in ICU units. We make use of the epidemiological bulletin published by the RKI in 2020, \cite{RKI_ICU_survival}, see Figure \ref{fig:Percentage_Outgoing} in the appendix. 
Finally, we impose the standard constraints, i.e. that both $s^I(t)$ and $s^R(t)$ as well as the spatial effects $h^I(\mbox{longitude}_d, \mbox{latitude}_d) $ and $h^R(\mbox{longitude}_d, \mbox{latitude}_d)$ integrate out to one. We refer to \citet{WOOD:2017} for more details.

\subsection{SEM algorithm}
Note that estimation of models (\ref{eq:model-in}) and (\ref{eq:model-out}) would be straightforward if we knew the data $I_{(t,d)}$ and $R_{(t,d)}$. This suggests making use of an EM algorithm by treating the estimation as a missing data problem. Note that we observe $\Delta_{(t,d)}$ from which we can "calculate" $I_{(t,d)}$ and $R_{(t,d)}$ . Given the data we have 
\begin{align}
    I_{(t,d)} = \Delta_{(t,d)} + R_{(t,d)}
\end{align}
with the additional constraints that both, $I_{(t,d)} \ge 0$ and $R_{(t,d)} \ge 0$. Hence, based on the data, we have the joint  probability model for incoming and released ICU patients: 
\begin{align}
\nonumber
  &   P( I_{(t,d)} = k , R_{(t,d)} = j | \Delta_{(t,d)} = \delta ) \\
  & 
  \label{eq:conv}
  \propto 
    \left\{ 
    \begin{tabular}{cl}  
   $ P( I_{(t,d)} = k)  \times P(R_{(t,d)}= k - \delta ) $ & \mbox{ for } $ j = k - \delta $ \mbox{ and }  $ k \ge \mbox{ max}( \delta , 0 )$ \\
    $0$ & \mbox{ otherwise }
    \end{tabular}
    \right. 
\end{align}
with $ P( I_{(t,d)} = k) $ and $P(R_{(t,d)}= k - \delta )$ resulting from the Poisson model (\ref{eq:model-in}) and (\ref{eq:model-out}), respectively. While model (\ref{eq:conv}) is a clumsy convolution model which does not simplify to an analytic form, simulation from the model is simple by just replacing the infinite pairs $k$ for $I_{(t,d)}$ and $k-\delta $ for $R_{(t,d)}$ by a set of finite pairs, such that the resulting cumulative probabilities are  approximately equal to one. To be specific, we have
\begin{align}
\nonumber P( I_{(t,d)} = k , R_{(t,d)} = j &| \Delta_{(t,d)} = \delta , \lambda^{I}, \lambda^{R})\\
 &=\lim_{K\to\infty}  
\frac{P_{\lambda^{I}}(I_{(t,d)} = k)P_{\lambda^{R}}(R_{(t,d)} = k-\delta)}{\sum_{i=1}^KP_{\lambda^{I}}(I_{(t,d)} = i)P_{\lambda^{R}}(R_{(t,d)} = i-\delta)}. \label{eq:jointprobexact}
\end{align}
We approximate this numerically by assuming that either $P_{\lambda^I}({I_(t,d)} = k)$, or $P_{\lambda^R}(R_{(t,d)}=k-\delta)$ is sufficiently close to zero at $k> 1000$ making the product of the two distributions sufficiently close to zero, such that the sum of probabilities for events $k> 1000$ can be ignored. This results in the finite approximation 
\begin{align}
\nonumber P( I_{(t,d)} = k , R_{(t,d)} = j &| \Delta_{(t,d)} = \delta , \lambda^{I}, \lambda^{R})\\
 &\approx \frac{P_{\lambda^{I}}(I_{(t,d)} = k)P_{\lambda^{R}}(R_{(t,d)} = k-\delta)}{\sum_{i=1}^{1000}P_{\lambda^{I}}(I_{(t,d)} = i)P_{\lambda^{R}}(R_{(t,d)} = i-\delta)}. \label{eq:jointprobapprx}
\end{align}

Numerically this is easily carried out and allows to simulate data pairs $( I_{(t,d)}^* , R_{(t,d)}^*)$ based on the current estimates of the intensities using (\ref{eq:jointprobapprx}) as an approximate version of (\ref{eq:conv}). This provides a stochastic `E'-step and leads to a full data set with (simulated) incoming and (simulated) released patients for all districts and all time points. With the resulting (simulated) full data set, we can now directly estimate the intensities in models (\ref{eq:model-in}) and (\ref{eq:model-out}), which in turn is conducted in the `M' step. The `M' step can be carried out by fitting two generalized additive Poisson models using standard software, see \cite{WOOD:2017}.

Iterating between the two steps gives a stochastic version of the EM algorithm. Each simulation step provides an estimate, and following the classical EM algorithm, we can easily see that on average, we increase the (marginal) likelihood in each step. 
The outline of which is sketched in Algorithm \ref{alg:EMalg}, see Appendix.

\subsection{Inference based on SEM}

Unlike the EM algorithm, where calculating the variance of the estimates is not straightforward, and one typically relies on Louis' formula \citep{10.2307/2345828}, the stochastic version allows to take the uncertainty due to the missing data into account. The derivation shows similarities to Rubin's formula for imputation, see \citet{rubin1976inference}. Let the parameter vector of linear and smooth functions, $\hat{\boldsymbol{\beta}}^{(*k)} = ( \hat{\boldsymbol{\beta}}^{I(*k)^T},\hat{\boldsymbol{\beta}}^{R(*k)^T})^T$, be the resulting estimate in the $k^{th}$ step of the SEM algorithm. We assume $k > k_0$, where $k_0 $ refers to the step when convergence can be guaranteed. The final estimate results through
\begin{align}
 \label{eq:EstimatorBeta}
    \hat{\boldsymbol{\beta}} = \frac{1}{K-k_0} \sum_{k=k_0+1}^{K} \hat{\boldsymbol{\beta}}^{(*k)}.
\end{align}
The variance is estimated via
\begin{align}
 \label{eq:VarianceRubin}
    \widehat{\mbox{Var}}(\hat{\boldsymbol{\beta}}) = 
  &\frac{1}{K-k_0}\sum_{k=k_0+1}^{K}\widehat{Var}(\hat{\boldsymbol{\beta}}^{(*k)})+ \frac{1+(K-k_0)^{-1}}{(K-k_0)-1}\sum_{j=k_0 + 1}^{K}
  (\hat{\boldsymbol{\beta}}^{(*k)}- \hat{\boldsymbol{\beta}}) (\hat{\boldsymbol{\beta}}^{(*k)}- \hat{\boldsymbol{\beta}}) ^T
\end{align}
where $\widehat{Var}(\hat{\boldsymbol{\beta}}^{(*k)})$ is the variance estimate in the $k$ iteration step, based on the imputed data set. The latter directly results through the applied fitting algorithm.

\section{Results}
A great advantage of our approach is that we can directly interpret the estimated association between the included covariates and the incoming patients and outgoing patients separately. To do so, we look at  covariates containing information on the infection rates for each of the three age groups and the weekday effects, which are included as a categorical variable using reference coding, with Friday as its reference level. The estimated coefficients and their standard deviation, calculated based on Rubin's formula, see Equation \ref{eq:VarianceRubin}, are provided in Table \ref{tab:Estimates}. 
We use the last three hundred runs to determine the coefficient estimates through their median, as well as their variance through the Equation (\ref{eq:VarianceRubin}). The estimates over the last 300 runs are shown through line plots in Figures \ref{fig:Coefficients_Incoming} and \ref{fig:Coefficients_Outgoing} in the Appendix for the incoming and outgoing patients, respectively.

First, we look at the association between our covariates and the number of incoming and outgoing patients, as seen in the middle and right column of the output table, Table \ref{tab:Estimates}.  Recall that the weekday effect is included in the model through a categorical variable, with Friday as its reference category. For the model estimating the number of incoming patients, keeping respectively all other variables constant, we can observe that there is an increased number of incoming patients on other weekdays, compared to Friday, whereas on the weekend, there is a decreased number of patients, compared to Friday.  
For  outgoing patients, the behaviour is slightly different. On Monday, Thursday, Saturday and especially Sunday, fewer patients are released compared to Friday. Conversely, Tuesday and Wednesday seem slightly increased. 

The number of incoming and outgoing patients is positively associated with the infection rates of all age groups. 
Notably, the strongest effect exists for the infection rate of 35 to 59-year-olds. This is interesting, bearing in mind that   60 to 69-year-olds are the predominant age group in intensive care. We should, however, not omit that there is strong collinearity between the infection rates themselves, as seen in Figure \ref{fig:Correlation} in the Appendix.  


\begin{table}[h]
\begin{center}\normalsize
\begin{tabular}[t]{l||c|c|c|c}
\hline
  & \multicolumn{2}{c}{Incoming} & \multicolumn{2}{c}{Outgoing}\\
  &Estimates & Std. Dev. &Estimates & Std. Dev.\\
\hline
Intercept & -2.28 & 0.10 & -6.41 & 0.12\\
Monday Effect & 0.12 & 0.05 & -0.21 & 0.06\\
Tuesday Effect & 0.14 & 0.05 & 0.03 & 0.06\\
Wednesday Effect & 0.13 & 0.05 & 0.02 & 0.06\\
Thursday Effect & 0.14 & 0.05 & -0.10 & 0.06\\
Saturday Effect & -0.02 & 0.05 & -0.09 & 0.06\\
Sunday Effect & -0.14 & 0.05 & -0.39 & 0.06\\
Infection 35-59 yo & 0.24 & 0.05 & 0.28 & 0.06\\
Infection 60-79 yo & 0.07 & 0.05 & 0.07 & 0.05\\
Infection 80+ yo & 0.11 & 0.02 & 0.10 & 0.02\\
\hline
\end{tabular}
\end{center}
\caption{Estimated coefficients and standard deviations presented on the level of incoming and outgoing patients. The estimates are the exponential of the median of the coefficient estimates from the 200$^{th}$ run to the 500$^{th}$ run of the EM algorithm.}
\label{tab:Estimates}
\end{table}

\begin{figure}[h]
    \centering
\includegraphics[width=0.9\textwidth, height= 200pt]{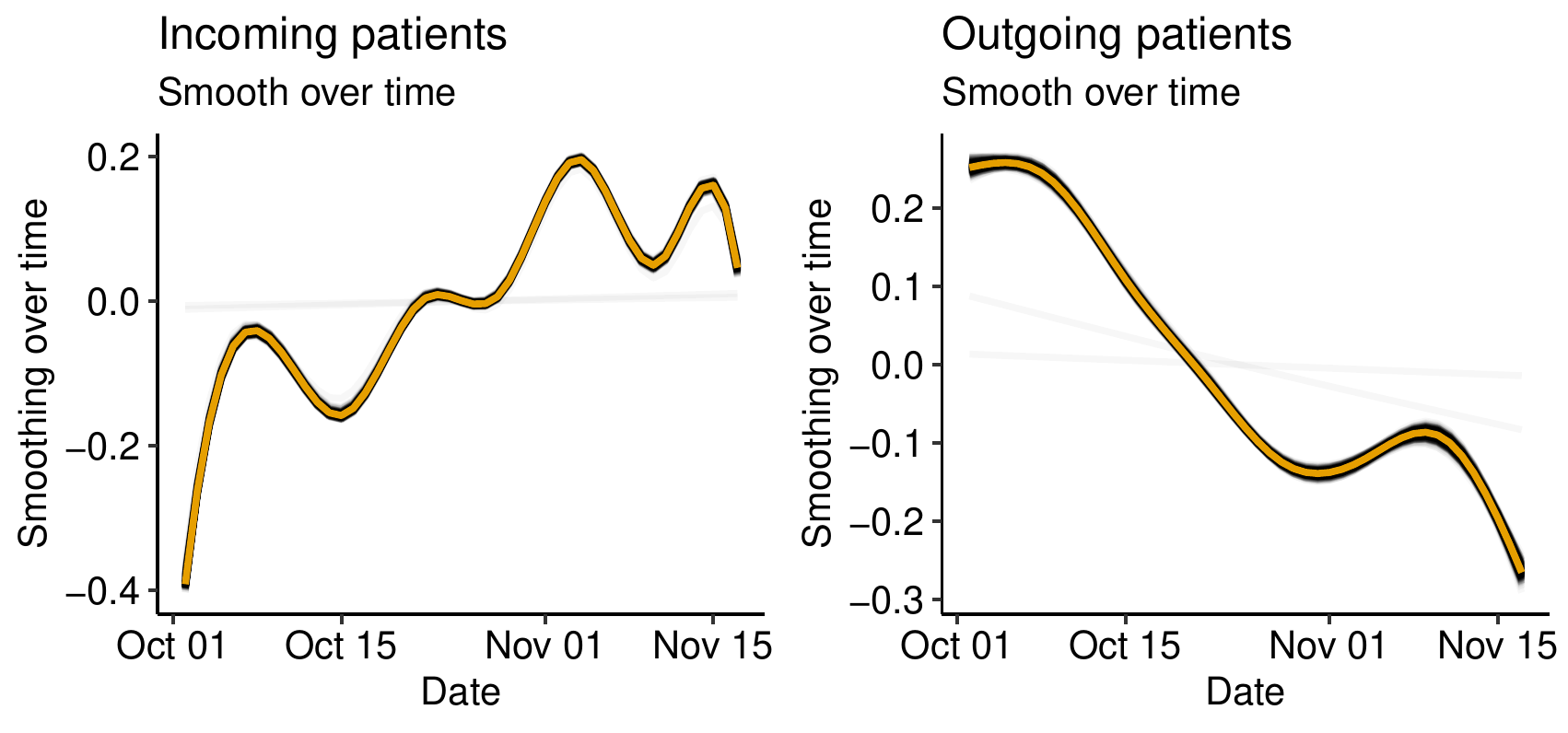}
   \caption{Estimated smooth functions of all runs, over time, rendered by the GAMs estimating the number of incoming patients (left hand side) and outgoing patients (right hand side).
   }
    \label{fig:Timesmooth}
\end{figure}

We also included smooth functions to estimate both the spatial-, and the temporal effects. They are included to pick up on additional spatial and temporal structural dependencies. 
Let us first  look at the smooth effects in time, as seen in Figure \ref{fig:Timesmooth}.
The averaged smooth function over time for incoming patients (left-hand side) is generally increasing.  Evidently, we can see some fluctuation and
there seems to be a fortnightly rhythm within the overall trend. Here we observe an increase in the number of incoming patients for the first seven days, then a decrease in the following seven days, followed by a subsequent increase, and so forth.   
In contrast, as shown on the  right-hand side of Figure \ref{fig:Timesmooth}, we see a general decrease in the number of outgoing patients without a biweekly rhythm. 

\begin{figure}[h]
    \centering
\includegraphics[width=0.7\textwidth]{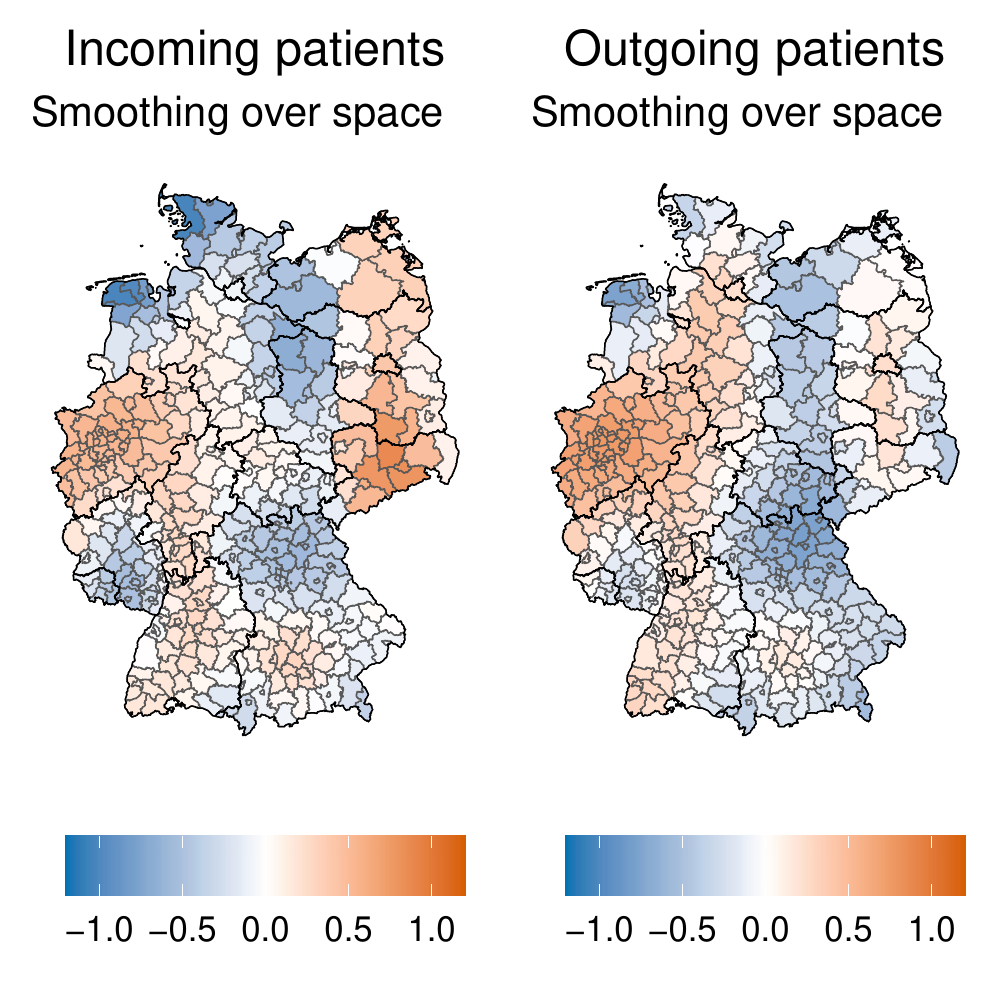}
   \caption{Estimated median smooth functions of the $200^{th}$ until the $500^{th}$ run over space rendered by the generalized additive models, estimating the number of incoming patients (left-hand side) and outgoing patients (right-hand side).
   }
    \label{fig:Spacesmooth}
\end{figure}

Finally, we look at the spatial effects  for the incoming patients, see  the left-hand side of Figure \ref{fig:Spacesmooth}, and for the outgoing patients, shown with the right-hand side of Figure \ref{fig:Spacesmooth}. There seems to be an increased level of incoming patients in Saxony (east Germany) and North Rhine-Westphalia (west Germany) and a slight increase around the larger cities of Germany (Frankfurt, Stuttgart and Munich, south and southwest of Germany). We observe a similar structure in the spatial smooth function in the model estimating the outgoing patients, except for the the strong increase around Saxony. Overall, we see clear spatial heterogeneity.

\begin{figure}[h]
    \centering
\includegraphics[width=0.8\textwidth, height= 250pt]{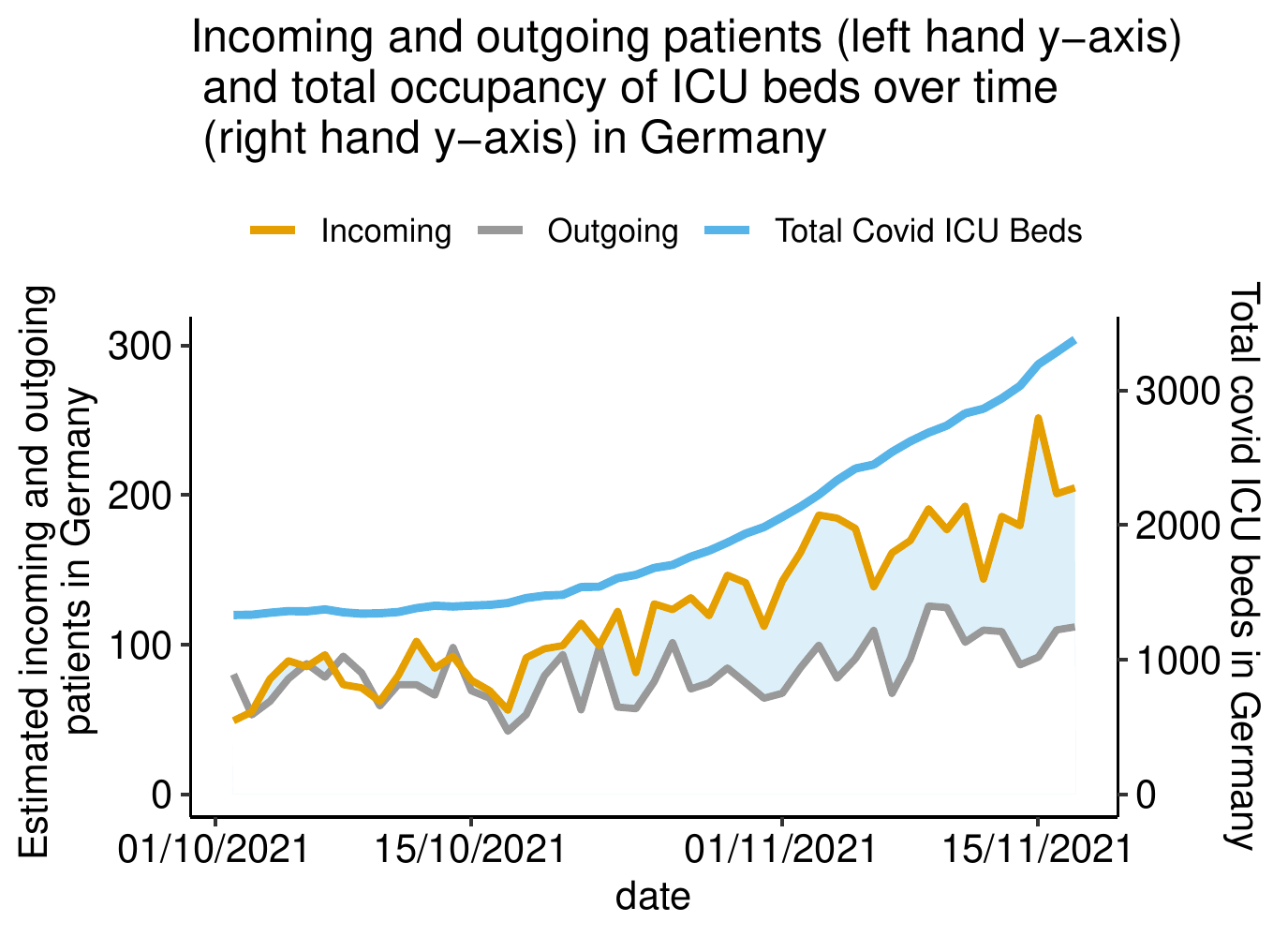}
   \caption{Estimated number of incoming and outgoing patients by date from the $1^{st}$ of October 2021 until the $18^{th}$ of November 2021, as well as the total number of COVID-19 patients in the ICUs of Germany.}
    \label{fig:Incoming_Outgoing}
\end{figure}

At last, we visualise in Figure \ref{fig:Incoming_Outgoing}  the estimated number of incoming and outgoing patients, summed up over the entirety of Germany, for the observed time frame.  The left-hand axis scales the number of incoming and outgoing patients, whereas the right-hand axis scales the number of overall ICU patients with COVID-19. We see that the model picks up the somewhat constant occupancy, from the $1^{st}$ of October 2021, until the $17^{th}$ of October 2021, in Germany's ICUs rather well, where the number of incoming and outgoing patients are estimated to be similar, if not equal. Thereafter, the number of ICU patients in the ICU increases, around this time, we also observe a higher estimated number of incoming patients than outgoing patients. It is not unusual for patients, especially the critically ill, to stay in the ICU for more than four weeks, making the divergence in estimation for the number of incoming patients and outgoing patients entirely plausible.  





\section{Simulation}

To investigate the modelling approach we chose a simple version to emulate the data used above. We use one covariate, randomly drawn from a normal distribution, whose mean and variance are taken from the observed mean and variance of the logged infection rates of `60-79' year-olds. We choose this age group as `60-69' year-olds are the predominant group in the German ICUs during the fourth wave, as described in Section 2. The coefficients for the simulation are chosen in a way such that the difference in the simulated incoming and outgoing patients is somewhat similar to the range of the difference in the observed incoming and outgoing patients, namely $(-24, 20)$ in the observed data. 

\begin{align}
 \label{eq:XSim}
X \sim N(1.978, 1.397)
\end{align}

\begin{figure}[h]
    \centering
\includegraphics[width=\textwidth]{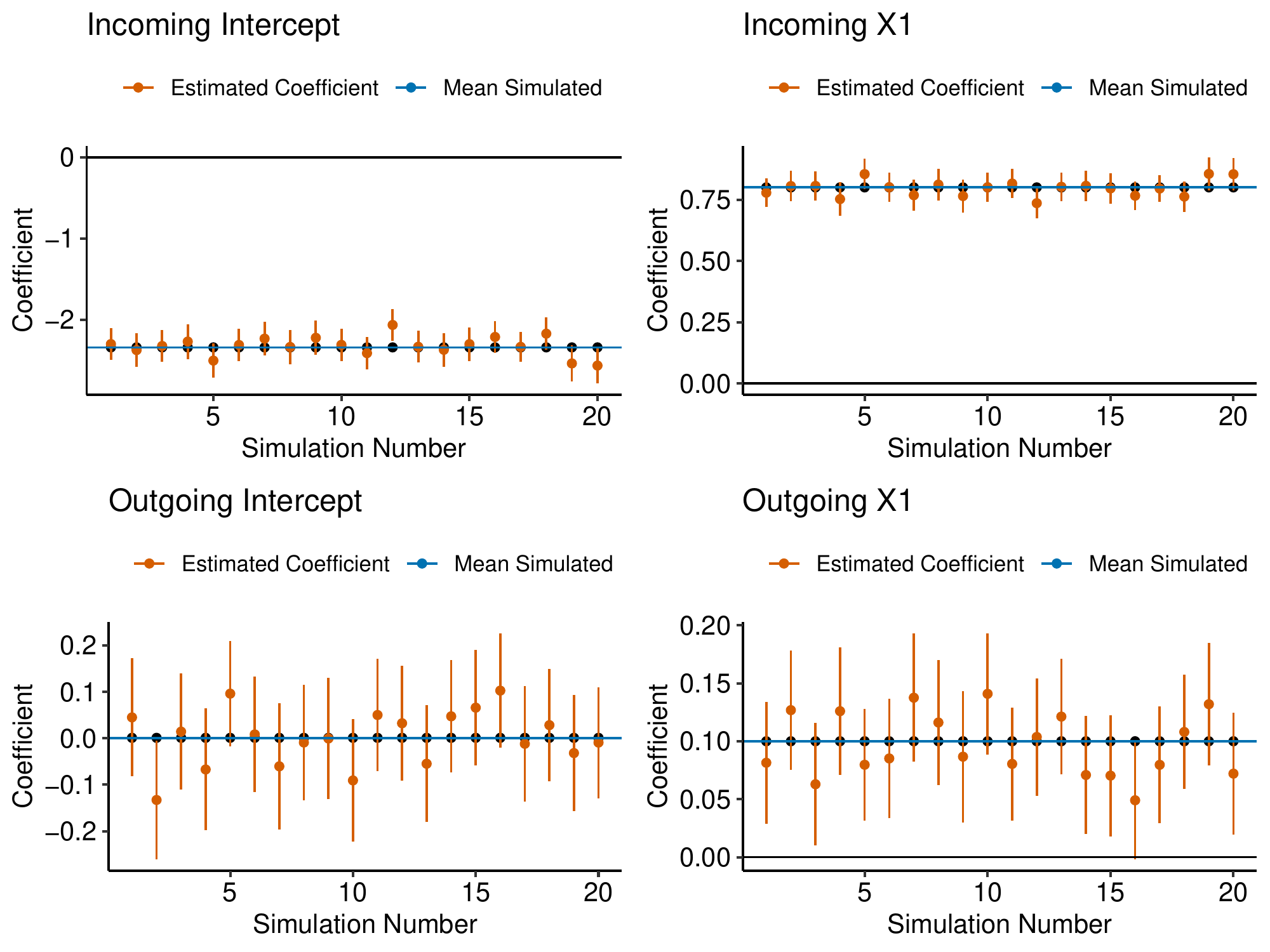}
   \caption{The estimated coefficients for twenty simulated data sets.}
    \label{fig:CoefficientsSim}
\end{figure}

The incoming and outgoing number of patients are then simulated as shown in equation \ref{eq:yinandyout}. The estimated coefficients in twenty runs are shown in Figure \ref{fig:CoefficientsSim}. 

\begin{align}
 \label{eq:yinandyout}
\nonumber I&\sim Poi(exp(\beta^{in}_0+ \beta^{out}_1 X))\\
R & \sim Poi(exp(\beta^{out}_0+ \beta^{out}_1 X+ log(I_{lag})))
\end{align}

We see for the confidence intervals of each of the coefficient estimates of each of the twenty runs include the real coefficient, except for the ``Incoming Intercept'' coefficient in the $12^{th}$ simulated data set. 
Overall, the simulation confirms that we are able to uncover incoming and outgoing patients from pure hospitalisations.

\begin{algorithm}
\caption{Simulation algorithm}\label{alg:Sim}
\begin{algorithmic}
\Require $n = 1000 + 1$ \Comment{Simulate a data set with 1000 entries, we need to start with $n=1000+1$. A seed is also set.} 
\State $X$ \Comment{Draw $X_c$ as a vector of $n$ entries from a Normal distribution with mean $1.978$ and variance $1.397$. This will be the covariate used to simulate the number of incoming and outgoing patients, $I^{sim}$ and $R^{sim}$. Let the design matrix, $X$, be a matrix of a $1 \times n$ dimensional one vector column bound to $X_c$.} 
\State $\vec{\beta}_{In}$ and $\vec{\beta}^{Out}$ \Comment{Set parameter vector association between responses and covariate. Here, $\vec{\beta}^{In}=(\beta^{In}_0, \beta^{In}_1)=(-2.340, 0.8)$ and $\vec{\beta}^{Out}=(\beta^{Out}_0, \beta^{Out}_1)=(0.001, 0.1)$} 
\State $I_{sim} \sim Poi(exp(\vec{\beta}^{In}X^T))$ \Comment{For each entry in $X$ respectively draw a number of incoming patients from a Poisson distribution, $Poi(exp(\vec{\beta}^{In}X^T))$.}
\State $R_{sim} \sim Poi(exp(\vec{\beta}^{Out}X_{Out}^T+log(I^{sim}_{Lag}+0.1)))$ \Comment{Let $X_{Out}$ be the last $n-1$ rows of $X$. Let $I^{sim}_{Lag}$ be the first $n-1$ entries of $I^{sim}$. Draw $n-1$ entries from the Poisson distribution, $Poi(exp(\vec{\beta}^{Out}X_{Out}^T+log(I^{sim}_{Lag}+0.1)))$.} 

\State \textbf{Return} \mbox{Simulated data} \Comment{Join the last $n-1$ entries of $I^{sim}$ with $R^{sim}$, calculate and join their difference $I^{sim}-R^{sim}=\delta^{sim}$. Further, join with the last $n-1$ entries of $X_c$.}
\end{algorithmic}
\end{algorithm}

\section{Concluding remarks}

Overall, in this application of the SEM, we are not only able to simulate unobserved data but also estimate the association between the weekday effect and the infection rates and the number of incoming and outgoing patients in a simple and intuitive manner. We achieve some insight into the estimated association between the infection rates and the number of incoming and outgoing patients. Namely, the driving force of the estimated number of incoming and outgoing patients seems to be the infection rates of `35-59' year-olds. Although we are not able to validate the predictions against the actual number of incoming and outgoing ICU patients, our findings seem to be mostly reasonable. Additionally, the SEM estimate the association of the simulated number of incoming and outgoing ICU patients and the simulated covariate, which are obviously observed measures in all but one simulation. In this situation, the SEM seems to be an appropriate application and allows us to gain a more complete picture of the COVID-19 pandemic, even when dealing with incomplete information.

 

\section*{Acknowledgements}
We acknowledge the support of the Deutsche Forschungsgemeinschaft (DFG, Project KA 1188/13-1). 

\section*{Declaration of conflicting interests}

The authors declared no potential conflicts of interest with respect to the research, authorship and/or publication of this article.


\bibliography{smj-template}

\newpage
\appendix
\section{Sketch of Algorithm}

\begin{algorithm}
\caption{EM algorithm}\label{alg:EMalg}
\begin{algorithmic}[h!]
\Require $\delta$ \Comment{$\delta$ is the observed difference.}
\State $\lambda^0_{In}$, $\lambda^0_{Out}$ \Comment{Choose starting values for the Poisson parameters of the number of incoming and outgoing patients, respectively.}
\State $P(I^0_{(t,d)}=k, R^0_{(t,d)}=k-\delta| \lambda^0_{In}, \lambda^0_{Out})\sim Poi_{\lambda^0_{In}}(k)Poi_{\lambda^0_{Out}}(k-\delta)$ 
\Comment{Simulate $I^0_{(t,d)}$ and $R^0_{(t,d)}$  from a joint probability distribution of two Poisson distributions, given the Poisson parameters previously chosen.}

\For{500 iterations- $i \in \left\{1, \dots, 500\right\}$}
    \State \textbf{M Step} \Comment{Estimate $\hat{\lambda}^i_{In}$ and $\hat{\lambda}^i_{Out}$ by using generalised additive models to estimate the association between the used covariates and the previously simulated number of incoming and outgoing patients, $I^{i-1}_{(t,d)}$ and $R^{i-1}_{(t,d)}$.}
    \State \textbf{`E'-step} \Comment{Simulate the number of incoming and outgoing patients through the joint probability distribution.}
    \State 
    $P(I^i_{(t,d)}=k, R^i_{(t,d)}=k-\delta| \lambda^i_{in}, \lambda^i_{out})\sim Poi_{\hat{\lambda}^i_{In}}(k)Poi_{\hat{\lambda}^i_{Out}}(k-\delta)$ 
\EndFor
\State \textbf{Return} \Comment{A list of estimated parameters (M-Step) and simulated number of incoming and outgoing patients (`E'-step) for each iteration.} 
\end{algorithmic}
\end{algorithm}

\newpage
\section{Additional Plots}
\begin{figure}[h!]
    \centering
    \includegraphics[width=0.9\textwidth]{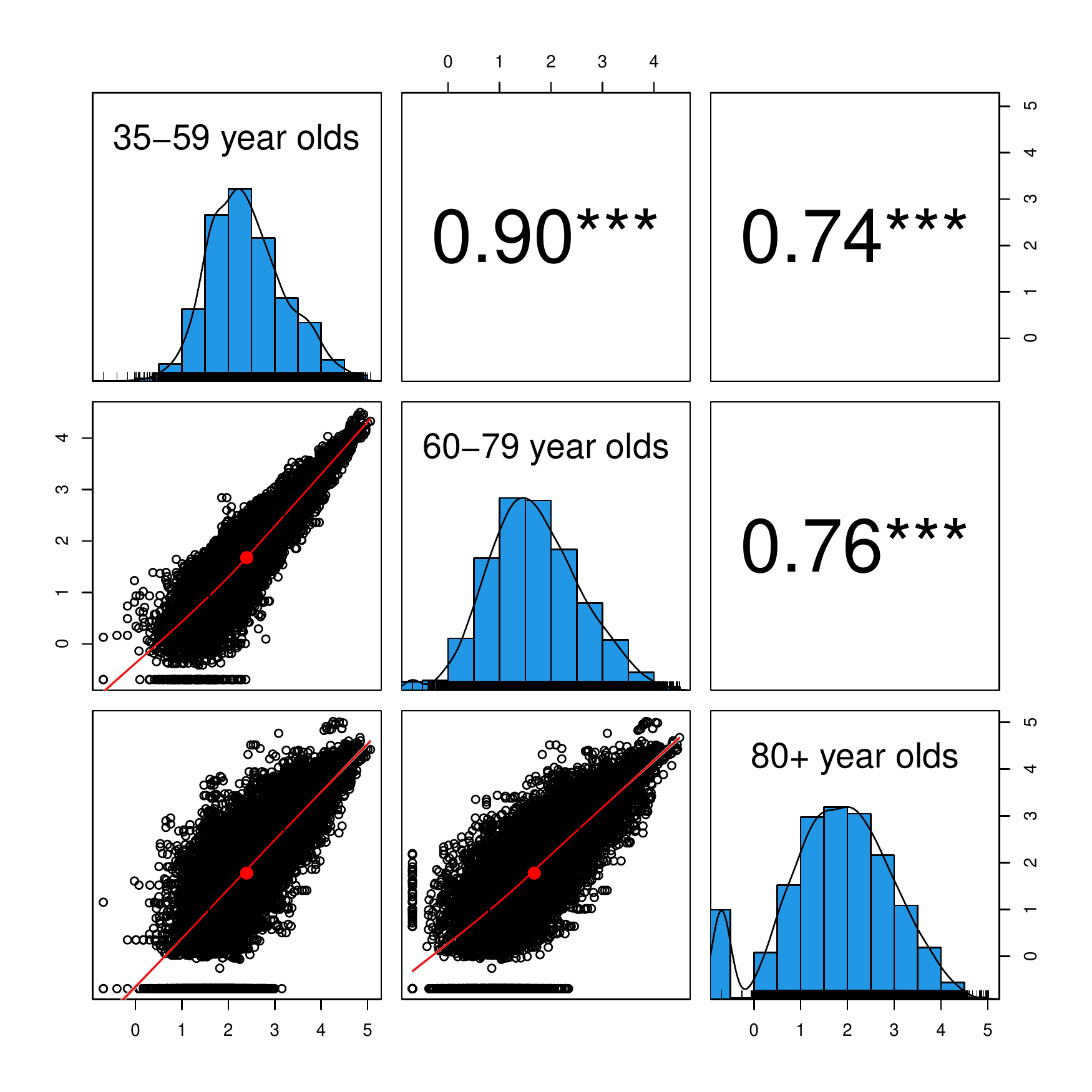}
    \caption{A plot outlining the linear correlation between the infection rates of `35-59' year-olds, `60-79' year-olds and `80+' year-olds. On the lower triangle, we see their respective scatter plots, along the diagonal, their distribution and on the upper triangle, their pairwise Pearson's correlation coefficients.}
    \label{fig:Correlation}
\end{figure}

\begin{figure}[h!]
    \centering
\includegraphics[width=0.9\textwidth]{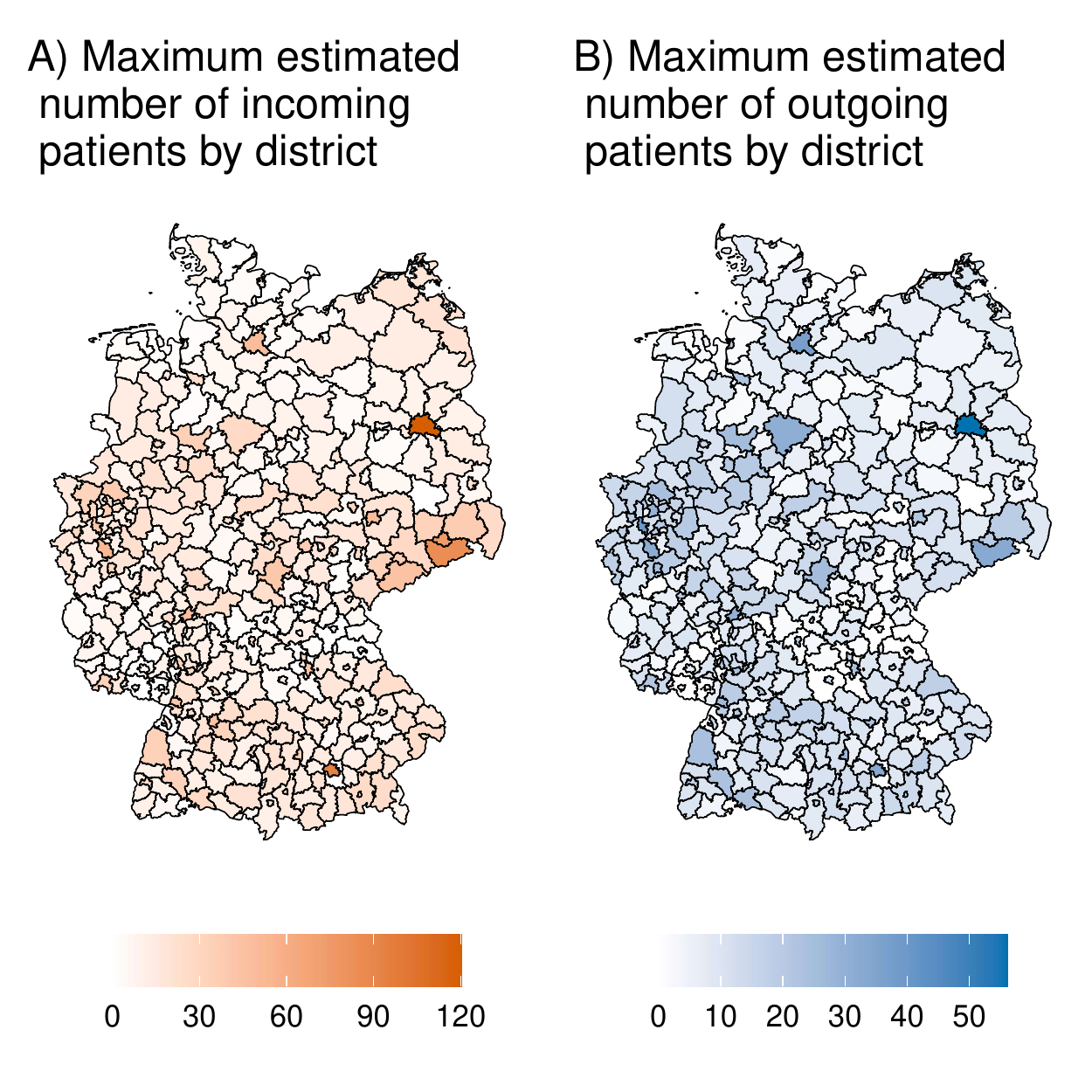}
   \caption{Estimated maximum number of incoming A) and outgoing patients B) per district from the $2^{nd}$ of October 2021 until the $17^{th}$ of November 2021.}
    \label{fig:Incoming_outgoing_max}
\end{figure}

\begin{figure}[h!]
    \centering
\includegraphics[width=0.5\textwidth]{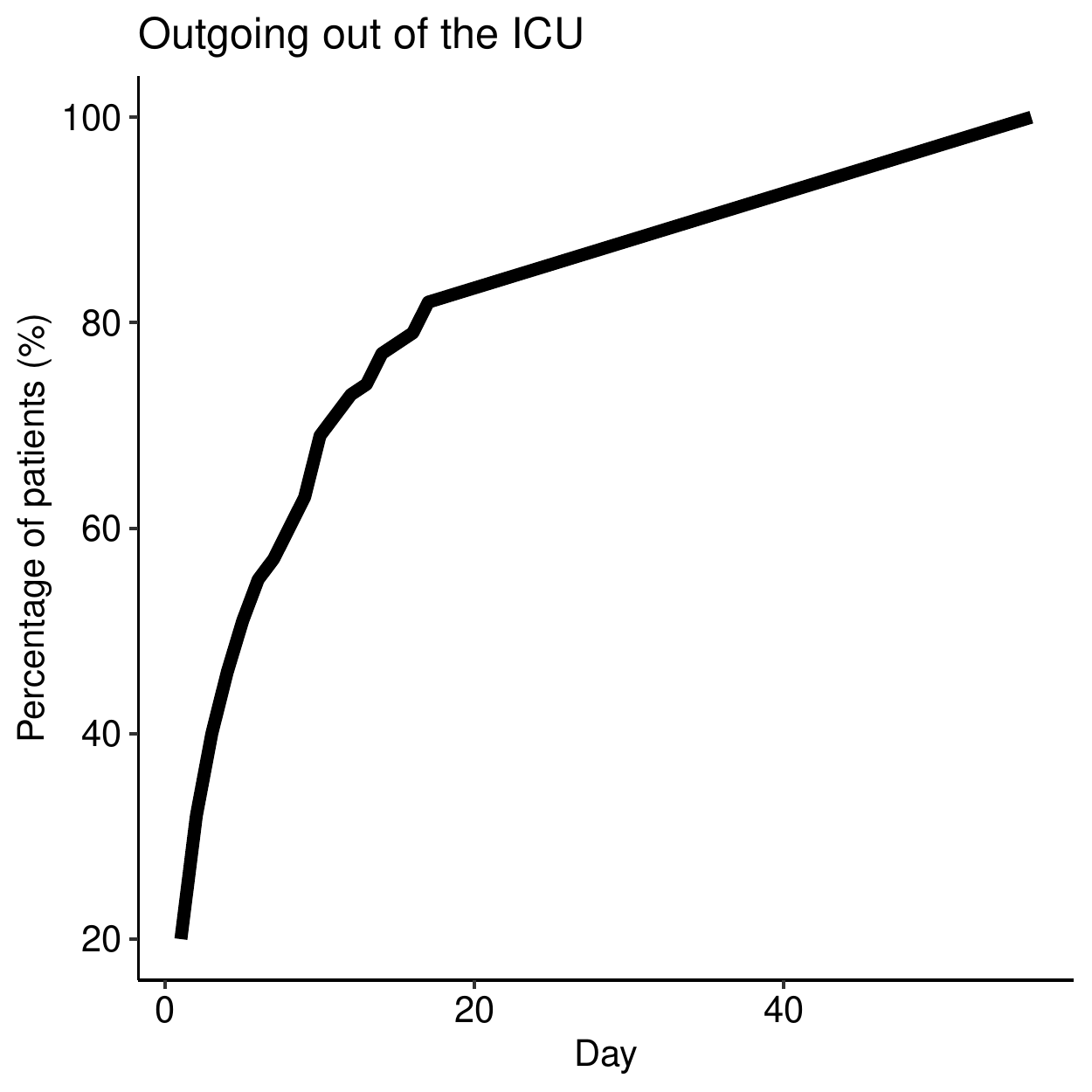}
   \caption{Percentage of outgoing ICU patients after the day of admission, as presented by \cite{Tolksdorf2020Eine}.}
    \label{fig:Percentage_Outgoing}
\end{figure}

\begin{figure}[h!]
    \centering
\includegraphics[width=0.7\textwidth]{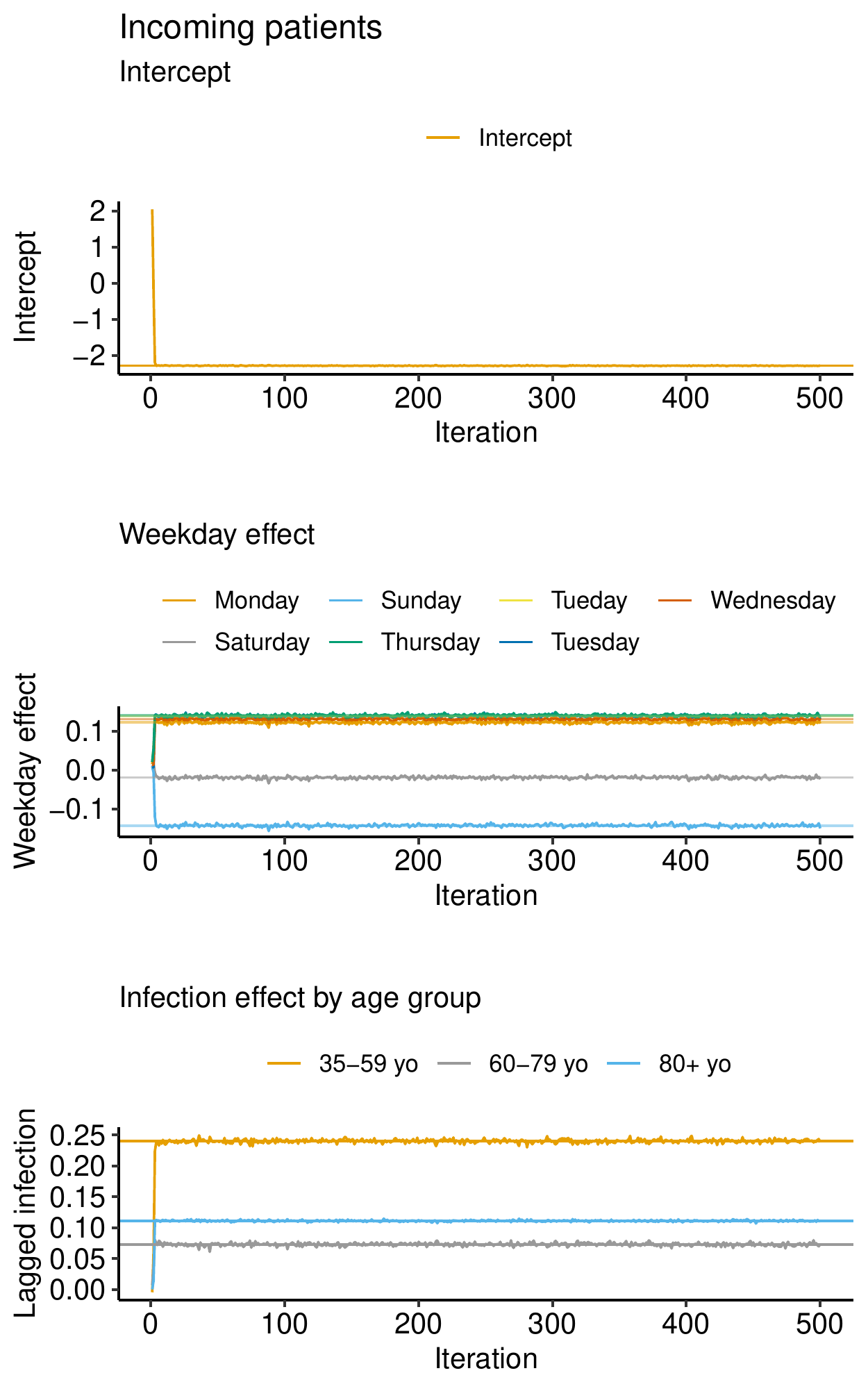}
   \caption{Coefficients estimated by the generalised additive models of the last three hundred runs of the EM algorithm of the incoming patients.}
    \label{fig:Coefficients_Incoming}
\end{figure}

\begin{figure}[h!]
    \centering
\includegraphics[width=0.7\textwidth]{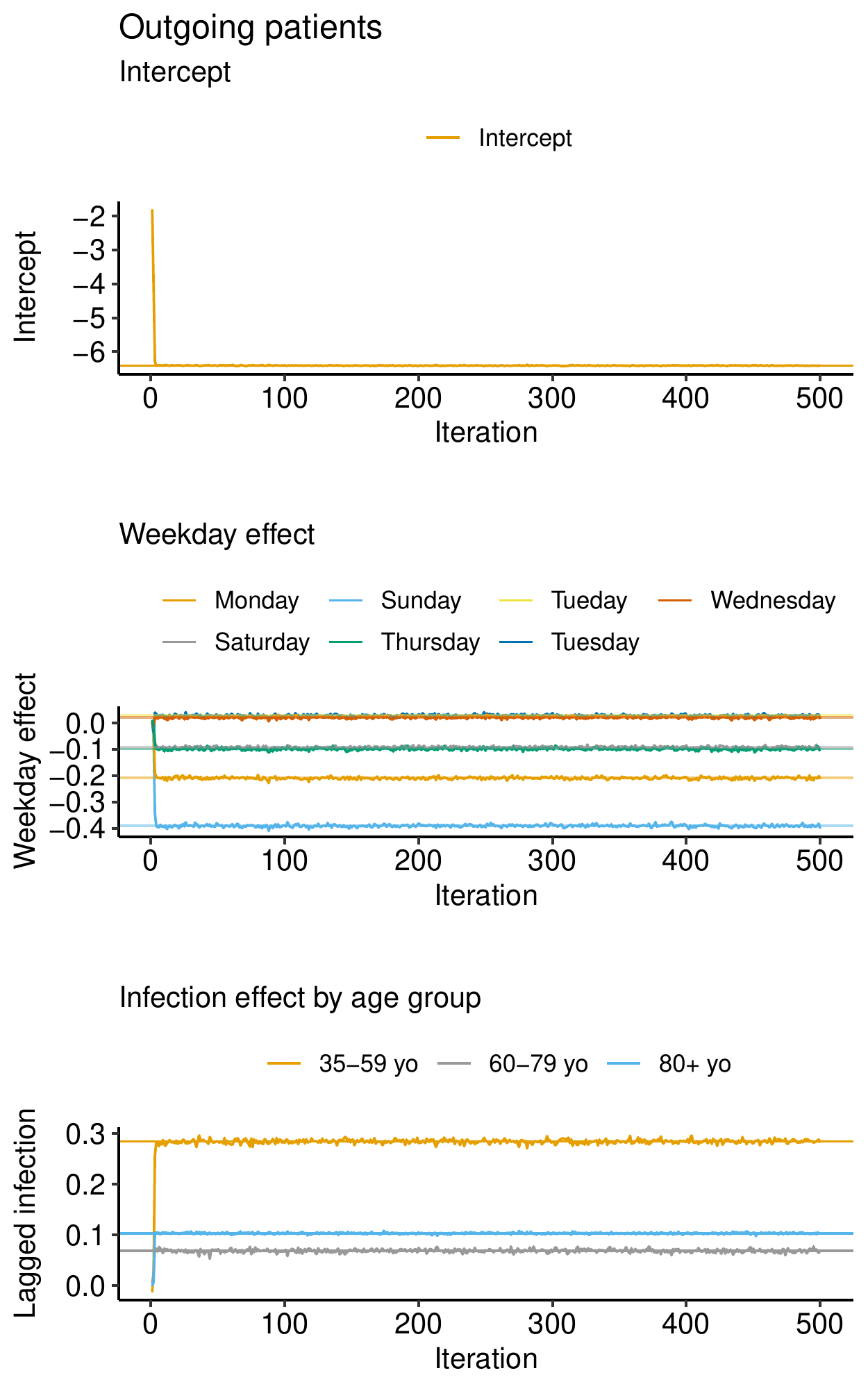}
   \caption{Coefficients estimated by the generalised additive models of the last three hundred runs of the EM algorithm of the outgoing patients.}
    \label{fig:Coefficients_Outgoing}
\end{figure}

\end{document}